# On the recovery of the time average of continuous and discrete time functions from their Laplace and z-transforms

*Emanuel Gluskin*[1,2] and *Shmuel Miller*[1]

[1]The ORT Braude Academic College, [2]The Kinneret College on the Sea of Galilee, and Ben-Gurion University of the Negev, Israel. gluskin@ee.bgu.ac.il

**Abstract**: The determination of the time averages of continuous functions, or discrete time sequences is important for various problems in physics and engineering, and the generalized final-value theorems of the Laplace and z-transforms, relevant to functions and sequences not having a limit at infinity, can be very helpful in this determination. In the present contribution, we complete the proofs of these theorems and extend them to more general time functions and sequences with a well-defined average. Besides formal proofs, some simple examples and heuristic and pedagogical comments on the physical nature of the limiting processes defining the averaging are given.

## 1. Introduction

### 1.1. General

In what follows, we use, for causal time functions $f(t)$ and sequences $f[n]$, one-directional Laplace and z- (Laurent) transforms:

$$F(s) = \int_0^\infty e^{-st} f(t)\, dt;$$

$$F(z) = \sum_{m=0}^\infty f[m] z^{-m} = f[0] + f[1]z^{-1} + f[2]z^{-2} + \dots .$$

We also use the notation $\Leftrightarrow$ to express a transformation relationship,

$$f(t) \Leftrightarrow F(s) \quad \text{and} \quad f[n] \Leftrightarrow F(z).$$

The following generalization of the Laplace transform final-value theorem for periodic and almost-periodic time functions was previously reported [1]:

$$\lim_{s \to 0}[sF(s)] = \langle f(t) \rangle, \qquad (1)$$

where <*f(t)*> is the time average of *f(t)* in the interval $[0,\infty)$. If as $t \to \infty$ the limit (final value) $f(\infty) = \lim f(t)$, exists, then the time average equals this value $\langle f(t) \rangle = f(\infty)$, and then (1) is reduced to the well-known version of the theorem, given in numerous textbooks.

This generalized theorem has found physical applications [2-4] where the ergodicity of a system makes it possible to reduce the ensemble averages to the time averages along the trajectories of the particles involved, making (1) relevant to the determination of some macroscopic system parameters.

The *z*-transform version of the generalization, related to sequences $f[n]$ (say, sampled values of a continuous physical function $f(t)|_{t=nT_s}$, $n$ an integer and $T_s$ the inter-sampling interval),

$$\lim_{z \to 1}[(z-1)F(z)] = \langle f[n] \rangle \qquad (2)$$

was introduced in [1] for periodic and almost-periodic functions without any detailed proofs, and was used subsequently in operational research applications in [5, 6].

In the discrete case as well, if $f[\infty] = \lim_{n \to \infty} f[n]$ exists, then $\langle f[n] \rangle = f[\infty]$, and we come to the classical version of (2), the only one found in the textbooks.

Thus, the proofs of (1) and (2) given in [1] are insufficiently general or complete, and the associated description in [2-7] of the applications of the generalized theorem is insufficiently theoretically supported and simple, which motivated the present research.

Following the circuit-applicative line of [1], and using the widely accepted system theory notations *x(t)* for input and *y(t)* for output of a linear system with impulse response *h(t)*, we come to the important formula (see Section 4)

$$\langle y(t) \rangle = H(0) \langle x(t) \rangle, \qquad (3)$$

where H(*s*) is the Laplace transform of the impulse response *h(t)*. This is the "system application" of (1).

In full analogy to (3), we also obtain,

$$\langle y[n] \rangle = H(1) \langle x[n] \rangle, \qquad (4)$$

where $H(z)$ is the transfer function (i.e. z-transform of the unit impulse response $h[n]$) of a system having $x[n]$ at its input and generating $y[n]$ at its output.

It is worth recalling that the z-transform is associated with the "discretized" Laplace transform according to:

$$z \leftrightarrow e^{sT_s} \qquad (5)$$

where $T_s$ is the *sampling* interval (period). This correspondence allows one to easily forecast some results from the continuous-time case to the discrete-time case.

In terms of the sequences, aiming, in particular, at some system applications, we shall employ only causal sequences satisfying the equality

$$x[n] = x[n]u[n]$$

where *u(n)* is the discrete unit step (the sampled Heaviside's function) that equals 1 for nonnegative *n*, and elsewhere is 0.

*1.2 . The simple periodic case*

The final-value theorem for *periodic continuous-time* functions is detailed in [1], but not proved in the discrete case. The generalized discrete time final value theorem for periodic signals, however, can be easily elaborated for the discrete sequences too, if one starts from following well known result.

Lemma 1:

*If x[n] is N-periodic, that is, the N-elements subsequence {x[0], x[1], … , x[N-1]} is periodically repeated, starting from $n = 0, N, 2N, ...$, then the z-transform of x[n] is given by*

$$X(z) = \frac{\sum_{m=0}^{N-1} x[m] z^{-m}}{1 - z^{-N}} . \tag{6}$$

Then, from (6) and the identity $(1 - z^{-N})^{-1} = z^N (z-1)^{-1} (1 + z + z^2 + ... z^{N-1})^{-1}$, we immediately obtain:

Theorem 1:

*For N-periodic x[n],*

$$\lim_{z \to 1}[(z-1)X(z)] = \frac{\sum_{m=0}^{N-1} x[m]}{N} = \langle x[n] \rangle_N . \tag{7}$$

Note that the average of a periodic function or sequence over a period equals its average over the whole infinite interval, which, for the discrete case, semantically can be written as

$$\langle x[n] \rangle_N = \langle x[n] \rangle_\infty = \langle x[n] \rangle . \tag{8}$$

An example for (7) can be the one-directional ("causal") cosine sequence

$$x[n] = (\cos n\omega_o) u(n) . \tag{10}$$

If we choose in (10) $\omega_o$ such that

$$\cos \omega_o = 1 , \tag{11}$$

which means $\omega_o = 2\pi k$, $k$ an integer, then $\cos n\omega_o = \cos nk2\pi = 1$. In this case $\langle x[n] \rangle = 1 = x[\infty]$, obviously.

If we choose $\omega_o$ such that

$$\cos \omega_o = -1 , \tag{11a}$$

which means $\omega_o = \pi + 2\pi k$, i.e. $n\omega_o = \pi n + 2\pi kn$, then

$$\cos n\omega_o = (-1)^n. \tag{12}$$

In this case, $x[\infty]$ does not exist, but $\langle x[n] \rangle = 0$.

Also for any other $\omega_o \neq 2\pi k$, $x[\infty]$ does not exist, but $\langle x[n] \rangle = 0$. Let us check this situation, using (2) and the well-known [8] z-transform of (10)

$$\frac{1 - z^{-1}\cos\omega_o}{1 - 2z^{-1}\cos\omega_o + z^{-2}}. \tag{13}$$

Using (13) in (2) one easily obtains 1 for $\cos\omega_o = 1$, and 0 for any $\cos\omega_o \neq 1$, as it has to be.

Using the same sequence (10), we can also demonstrate the effectiveness of (7). For this, consider a system having the impulse response

$$h[n] = \alpha^n u[n], \quad \alpha < 1, \tag{14}$$

and having at its input

$$x[n] = (\cos n\omega_o)u(n).$$

Using that for (27)

$$H(z) = \frac{1}{1 - \alpha z^{-1}},$$

we obtain (7) using (4) as

$$<y(n)> = \frac{1}{1-\alpha} <\cos n\omega_o>,$$

from which for $\cos\omega_o \neq 1$ we obtain $<y[n]> = 0$, and for $\cos\omega_o = 1$,

$$<y[n]> = \frac{1}{1-\alpha}.$$

Thus, the use of (7) is very simple when it is clear a priori what $<x[n]>$ is.

### *1.3. A comment on almost-periodic functions*

The result (7) for a periodic sequence is correct for *any* however large a period $N$, and similarly for *T*-periodic time-continuous functions [1]. That is

$$\lim_{s \to 0}[sF_T(s)] = \frac{\int_0^T f_T(t)\, dt}{T}$$

is correct for however large a period *T*. Considering this, a composition of periodic functions having an irrational ratio of the periods and the linearity of the Laplace transform were used in [1] to extend the applications of (1) from periodic to some *almost-periodic* functions having an infinite period. Almost periodic discrete sequences, with equidistant sampling, are also relevant to the point.

However (see Section 2), the class of non-periodic functions relevant for the generalized theorems is much wider than the class of almost-periodic functions or sequences. In fact, according to the physical meaning of the limit $T \to \infty$, or $N \to \infty$, if the period $N$ in (7) becomes unlimited and the average $\langle x[n] \rangle$ exists, then (2) must be correct for a non-periodic sequence as well. Thus, the formalization for a general application framework, having been used in [2-6] on some intuitive basis, and without proof, is established here.

## 2. The extended final-value theorems

### 2.1. *Continuous time functions*

Theorem 2:

$$\lim_{s \to 0}[sF(s)] = \langle f(t) \rangle \qquad (15)$$

for a Laplace-transform pair $f(t) \Leftrightarrow F(s)$ for which $<f(t)>$ exists and

$$\lim_{s \to 0} \int_s^\infty \frac{F(\zeta)}{\zeta} d\zeta = \infty \ .$$

Proof:

It is suitable to start from the formula [8]:

$$\int_s^\infty \Psi(\zeta) d\zeta = \int_0^\infty e^{-st} \frac{\psi(t)}{t} dt \ . \qquad (16)$$

Letting in (16)

$$\psi(t) = \int_0^t f(\lambda) d\lambda \ ,$$

and using also the transform pair

$$\int_0^t f(\lambda) d\lambda \Leftrightarrow \frac{F(s)}{s},$$

we have for the running time average $<f>_t$ :

$$<f>_t \ = \ \frac{1}{t}\int_0^t f(\lambda) d\lambda \ = \ \frac{\psi(t)}{t} \ \Leftrightarrow \ \int_s^\infty \frac{F(\zeta)}{\zeta} d\zeta \ . \qquad (17)$$

Since, furthermore, we assume that

$$\lim_{t \to \infty} <f>_t = <f>$$

*exists*, we have, using (17), in the classical (that of the textbooks) final-value-theorem applied to the *time function* $<f>_t = \psi(t)/t$:

$$\lim_{s \to 0} [s \int_s^\infty \frac{F(\zeta)}{\zeta} d\zeta] = <f> . \qquad (18)$$

It remains to use L'Hospital's rule, obtaining in the left-hand side of (18)

$$\lim_{s \to 0} [s \int_s^\infty \frac{F(\zeta)}{\zeta} d\zeta] = \lim_{s \to 0} \left[ \frac{\int_s^\infty \frac{F(\zeta)}{\zeta} d\zeta}{\frac{1}{s}} \right]$$

$$= \lim_{s \to 0} \left[ \frac{-\frac{F(s)}{s}}{-\frac{1}{s^2}} \right] = \lim_{s \to 0} [sF(s)] .$$

That is, (18) yields

$$\lim_{s \to 0} [sF(s)] = <f>$$

establishing (15).

Observe that the classical final value theorem is now applied not to *f(t)* (or *f[n]*) by itself, but to the "averaged" *time-function* $<f(\lambda)>_t$ that converges to some numerical value $<f>_\infty$ (or $<f>$), i.e. we come to the simple formula that interests us, employing a function (the running average) which is formally more complicated than *f(t)*, but converging.

### *2.2. Discrete time sequences*

Theorem 3:

$$\lim_{z \to 1}[(z-1)F(z)] = \langle f[n] \rangle \qquad (19)$$

for a z-transform pair $f[n] \Leftrightarrow F(z)$ for which $\langle f[n] \rangle$ exists and

$$\lim_{z \to 1} \int_z^\infty \frac{1}{\xi - 1} X(\xi) d\xi = \infty .$$

Proof:

The following proof parallels that of the above continuous time case. Let $X(z)$ denote the z-transform of the causal sequence $x[n]$. First of all, we observe that since the z-transform of the unit step sequence $u[n]$ is $(1-z^{-1})^{-1}$, $|z|>1$, the z-transform of

$$\sum_{k=0}^{n} x[k] = \sum_{k=0}^{\infty} x[k]u[n-k] = x[n]*u[n]$$

(the convolution) is

$$\frac{X(z)}{1-z^{-1}}, \quad |z|>1. \tag{20}$$

Lemma 2: *The z-transform of the running time average,*

$$g[n] = \frac{1}{n} \sum_{k=0}^{n} x[k], \tag{21}$$

is

$$G(z) = \int_{z}^{\infty} \frac{X(\xi)}{\xi-1} d\xi. \tag{22}$$

Proof: Since $-z\,d(z^{-n})/dz = nz^{-n}$, we have the transform pair

$$n\,x[n] \Leftrightarrow -z\frac{dX(z)}{dz}.$$

Thus, for

$$y[n] \equiv ng[n] = \sum_{k=0}^{n} x[k]$$

we have

$$Y(z) = -z\frac{dG(z)}{dz},$$

(where $g[n] \Leftrightarrow G[z]$), or

$$G(z) = \int_{z}^{\infty} \xi^{-1} Y(\xi) d\xi.$$

(The boundaries of the integration are chosen for convergence reasons.) Using here (20), we obtain

$$G(z) = \int_{z}^{\infty} \xi^{-1} \frac{1}{1-\xi^{-1}} X(\xi) d\xi = \int_{z}^{\infty} \frac{X(\xi)}{\xi-1} d\xi,$$

i.e. (22).

Assume now that the running average converges, i.e. that all the poles of $G(z)$ are inside the unit circle, and apply the (classical) final value theorem to $g[n]$, namely $g[\infty] = \lim_{z \to 1}(z-1)G(z)$, or

$$\langle x \rangle = \lim_{n \to \infty} \frac{1}{n} \sum_{k=0}^{n} x[k] = \lim_{z \to 1}(z-1)G(z) \ .$$

Employing (22), this can be rewritten as

$$\langle x \rangle = \lim_{n \to \infty} \frac{1}{n} \sum_{k=0}^{n} x[k] = \lim_{z \to 1}(z-1)\int_{z}^{\infty} \frac{X(\xi)}{\xi-1} d\xi = \lim_{z \to 1} \frac{\int_{z}^{\infty} \frac{1}{\xi-1} X(\xi) d\xi}{\frac{1}{z-1}},$$

and L'Hôpital's rule gives:

$$\langle x \rangle = \lim_{n \to \infty} \frac{1}{n} \sum_{k=0}^{n} x[k] = \lim_{z \to 1} \frac{-\frac{1}{z-1} X(z)}{-\frac{1}{(z-1)^2}} = \lim_{z \to 1}(z-1)X(z) \ , \quad (23)$$

and (19) is proved.

One notes that the above general result may be hinted from (7) with $N \to \infty$ in (7), i.e. we can "simply" transfer in (7) to some general *nonperiodic sequences,* obtaining

$$\lim_{z \to 1}[(z-1)X(z)] = \langle x[n] \rangle.$$

The appearing possibility of using (7) with an infinite period is a heuristically interesting point, and the following related physical argument, completing the formal derivations, should not be missed, in particular by a teacher.

### 3. Some physical interpretations of the formal proofs

The processes of $T \to \infty$, or $N \to \infty$, appearing in our problem, belong to the broad class of limit processes in mathematical physics, for which *empirical criteria* introduce the mathematical measures, thus leading to engineering understanding of the processes.

As a classical example, when the Fourier integral is derived as a limit ($T \to \infty$) of the Fourier series by taking periodic functions with larger and larger periods, it is not clear *when* the discrete spectrum of the Fourier series becomes the continuous one, i.e. the integral is obtained. It is said in [9, p. 433] that it is *very difficult* to justify this transfer by $T \to \infty$ to the Fourier integral, and in [9], as well as in [10] and some other respectable high-level mathematical textbooks, the transfer $T \to \infty$ is avoided, and instead it is directly shown that the "Fourier integral formula" (the inverse transform) indeed gives the initial time-function being transformed.

The difficulty in the derivation of the Fourier integral immediately disappears if one faces the physical situation. Since frequencies are always measured with some finite available precision, denoting the certain, *known* error of the measurement as $\delta\omega$, one sees that $T$ becomes "infinite" (i.e. the frequency spectrum "becomes continuous") already *when $T^{-1} \sim \Delta\omega < \delta\omega$, and the precise transfer $\Delta\omega \to 0$ (i.e. $T \to \infty$) used in the consideration of the continuous frequency spectrum, is simply not needed.* Thus, the criterion for the "infinite $T$", needed for the transfer to the Fourier integral, depends on the concrete physical conditions.

Similarly, *if the period of x[n] is larger than the total duration of the observation, then we can consider the periodic sequences as non-periodic.* The only essential distinction with respect to the continuous case is that while increasing the period of the sequence, we preserve the finite sampling period.

## 4. Applications to system theory

The possible system may be any linear circuit or control system, or a spatially distributed discrete system, like the purely resistive ladder and grid considered in [11], or a numerical linear "system" (or algorithm) that models in a computer a physical system or process. That is, application areas of the following result may be quite broad.

Theorem 4:

*If the sequence x[n] at the input of a linear time-invariant (i.e. having fixed parameters) stable system, characterized by a transfer function H(z), has an average, <x[n]>, then for the output sequence y[n]*

$$< y[n] > = H(1) < x[n] > . \qquad (24)$$

Proof:

Since the response $y[n] = (h*x)[n]$ becomes in the z-domain

$$Y(z) = H(z)X(z) , \qquad (25)$$

we obtain, using (2) that

$$\begin{aligned}< y[n] > &= \lim_{z \to 1}[(z-1)Y(z)] = \lim_{z \to 1}[(z-1)H(z)X(z)] \\ &= H(1)\lim_{z \to 1}[(z-1)X(z)] = H(1)\langle x[n] \rangle .\end{aligned} \qquad (26)$$

If $x(\infty)$ exists, then (24) can be also expressed as

$$y[\infty] = H(1) x[\infty] . \qquad (27)$$

In the time-continuous analogy,

$$< y(t) > = H(0) < x(t) > \qquad (24a)$$

and if $x(\infty)$ exists, then

$$y(\infty) = H(0) x(\infty) . \qquad (27a)$$

See also [1] where (27a) is proved, but with the limitations related to the limitations on the class of functions involved.

## 5. The role of asymptotic sequences

Disregarding whether or not a sequence $x[n]$ has a final value, i.e. a horizontal asymptote, if the *average* $<x[n]>$ exists it is defined only by the *asymptotic behavior of $x[n]$*.

For instance, if $x[n]$ is given as

$$x[n] = (1 + x_1[n])x_2[n], \qquad (28)$$

where

$$x_1[n] \to 0, \text{ as } n \to \infty,$$

then $x_2[n]$ is the asymptotic sequence for $x[n]$, and

$$< x[n] > = < x_2[n] > .$$

This means that very different sequences may have the same asymptotic behavior and thus the same average value.

Finally, a comment on the averages of a product of sequences follows invoking the known formula (e.g. [8], the integration path is inside ROC)

$$f[n]g[n] \Leftrightarrow \frac{1}{2\pi j}\oint F(\zeta)G(\frac{z}{\zeta})d\zeta = \frac{1}{2\pi j}\oint F(\frac{z}{\zeta})G(\zeta)d\zeta ,$$

and (2) to yield

$$< f[n]g[n] > = \lim_{z \to 1} (z-1) \frac{1}{2\pi j} \oint F(\zeta)G(\frac{z}{\zeta})d\zeta . \qquad (29)$$

If one of the sequences, here $g[n]$, is such that the limit may be moved into the integral, that is, we can assume that

$$\Psi(\zeta) = \lim_{z \to 1} (z-1)G(\frac{z}{\zeta})$$

*exists*, then (29), i.e. the calculation of the product-sequence's time average is simplified to

$$< f[n]g[n] > = \frac{1}{2\pi j} \oint F(\zeta)\Psi(\zeta)d\zeta .$$

## 6. Conclusions and final remarks

Generalizations of the final value theorem in the *s*-domain and *z*-domain have been considered. These generalizations relate to the *average* of the function or sequence, and not necessarily to its final value which may not exist. The presented material completes the traditional use of the s-transform and *z*-transform in the textbooks, and better supports the applications of the generalized theorems in [2-6].

The possibility of the transfer from the periodic sequence directly to a non-periodic one, is further and more deeply discussed here, in terms of both domains. It is made clear that the relevant functions need not be almost periodic, i.e. both (1) and (2) can be freely applied to physical processes, as was previously done without the proofs, in the application oriented works [2-6].

In the scientific regard, one should try to "translate" to the domain of the discrete variable *z* such approaches as, e.g. (see Section 3.2 of reference [7]) the application of the generalized theorem to the interesting field of "irrational" functions that include '*t*' (or '*n*') in a fractional degree. The recent work [12] enhances the results of [5-7] related to operational research; see also a recent book [13] and references therein, relevant to the irrational functions, and the use of the generalized theorem in [14].

Last but not least, work [15] attempts to further generalize the final-value theorems to the cases where the limiting values for $f(t)$ and for its running-average time-function, $f_1(t) = <f(\lambda)>_t$, do not exist. In these cases, repeated averaging, in the running-time sense, may be required until an average of a time function in $0 < t < \infty$ exists. As a remarkable point, it appears [15] that the respective limits in the *s* or *z* domain equal those in (1) or (2). However, it is not trivial to find applications with functions for which repeated averaging divergence occurs.